\documentclass[a4paper]{jpconf}
\usepackage{graphicx}
\usepackage{float} 
\usepackage{amssymb}
\usepackage{cite} 
\usepackage{hyperref}
\hypersetup{linkcolor=blue,citecolor=blue,filecolor=black,urlcolor=blue}
\newcommand{\pT} {\ensuremath{p_{\mathrm{T}}}}
\begin{document}
\title{Measurement of the rapidity-even dipolar flow in Pb-Pb collisions with the ATLAS detector}

\author{Jiangyong Jia on behalf of the ATLAS Collaboration}
\address{Chemistry Department, Stony Brook University, Stony Brook, NY 11794, USA}
\address{Physics Department, Brookhaven National Laboratory, Upton, NY 11796, USA}
\ead{jjia@bnl.gov}

\begin{abstract}
The rapidity-even dipolar flow $v_1$ associated with dipole asymmetry in the initial geometry is measured over a broad range in transverse momentum 0.5 GeV$<\pT<$9 GeV, and centrality (0-50)\% in Pb-Pb collisions at $\sqrt{s_{_{\mathrm{NN}}}}=2.76$ TeV, recorded by the ATLAS experiment at the LHC. The $v_1$ coefficient is determined via a two-component fit of the first order Fourier coefficient $v_{1,1}=\langle\cos \Delta\phi \rangle$ of two-particle correlations in azimuthal angle $\Delta\phi=\phi_{\mathrm{a}}-\phi_{\mathrm{b}}$ as a function of $\pT^{\mathrm a}$ and $\pT^{\mathrm b}$. This fit is motivated by the finding that the $\pT$ dependence of $v_{1,1}(\pT^{\mathrm a},\pT^{\mathrm b})$ data are consistent with the combined contributions from a rapidity-even $v_1$ and global momentum conservation. The magnitude of the extracted momentum conservation component suggests that the system conserving momentum involves only a subset of the event (spanning about 3 units in $\eta$ in central collisions). The extracted $v_1$ is observed to cross zero at $\pT\approx1.0$ GeV, reaches a maximum at 4--5 GeV with a value comparable to that for $v_3$, and decreases at higher $\pT$. Interestingly, the magnitude of $v_1$ at high $\pT$ exceeds the value of the $v_3$ in all centrality interval and exceeds the value of $v_2$ in central collisions. This behavior suggests that the path-length dependence of energy loss and initial dipole asymmetry from fluctuations corroborate to produce a large dipolar anisotropy for high $\pT$ hadrons, making the $v_1$ a valuable probe for studying the jet quenching phenomena. 
\end{abstract}

Recently, the measurements of harmonic flow coefficients $v_n$ have provided valuable insights into the properties of the hot and dense matter created in heavy ion collisions at the Relativistic Heavy Ion Collider (RHIC) and the Large Hadron Collider (LHC). These coefficients are obtained from a Fourier expansion of the particle production in azimuthal angle $\phi$:
\begin{eqnarray}
\label{eq:1}
\frac{dN}{d\phi}\propto1+2\sum_{n=1}^{\infty}v_{n}\cos n(\phi-\Phi_{n})
\end{eqnarray}
where $\Phi_n$ represents the phase of $v_n$. These $v_n$ coefficients are closely related to the Fourier coefficients $v_{n,n}$ of the two-particle correlation (2PC) in relative azimuthal angle $\Delta\phi=\phi_{\mathrm{a}}-\phi_{\mathrm{b}}$:
\begin{eqnarray}
\label{eq:2}
\frac{dN^{pairs}}{d\Delta\phi}\propto1+2\sum_{n=1}^{\infty}v_{n,n}\cos n(\Delta\phi),
\end{eqnarray}
which in general contains contributions from both harmonic flow $v_n$ and other non-flow correlations:
\begin{eqnarray}
\label{eq:3}
v_{n,n}(\pT^{\mathrm a},\pT^{\mathrm b})  = v_n(\pT^{\mathrm a})v_n(\pT^{\mathrm b})+{\mbox{non-flow}}
\end{eqnarray}
The $v_2$ coefficient, known to be associated with the elliptic shape of the overlap region, dominates in the non-central collisions. However, large $v_n$ values have also been measured for $n>2$~~\cite{Adare:2011tg,Aamodt:2011by,CMS:2012wg,Aad:2012bu}, which have been interpreted as the hydrodynamic response to the fluctuations of the matter in the initial states. The exotic ridge and Mach-cone like structures first observed at RHIC in the 2PC in $\Delta\phi$ and $\Delta\eta$~\cite{Adare:2008cqb,Abelev:2009qa,Alver:2009id} are naturally explained by the sum of these harmonics. The main sources of non-flow correlations include contributions from resonance decays, Bose-Einstein correlation, jets, and the global momentum conservation (GMC), a source that is very important for the $n=1$ term. This proceedings focuses on the $v_1$ results extracted from $v_{1,1}$ by taking into account the contribution of GMC in Eq.~\ref{eq:3}, using a method similar to Ref.~\cite{Retinskaya:2012ky}. The main results have been published in Ref.~\cite{Aad:2012bu}.

\begin{figure}[h]
\includegraphics[width=1\linewidth]{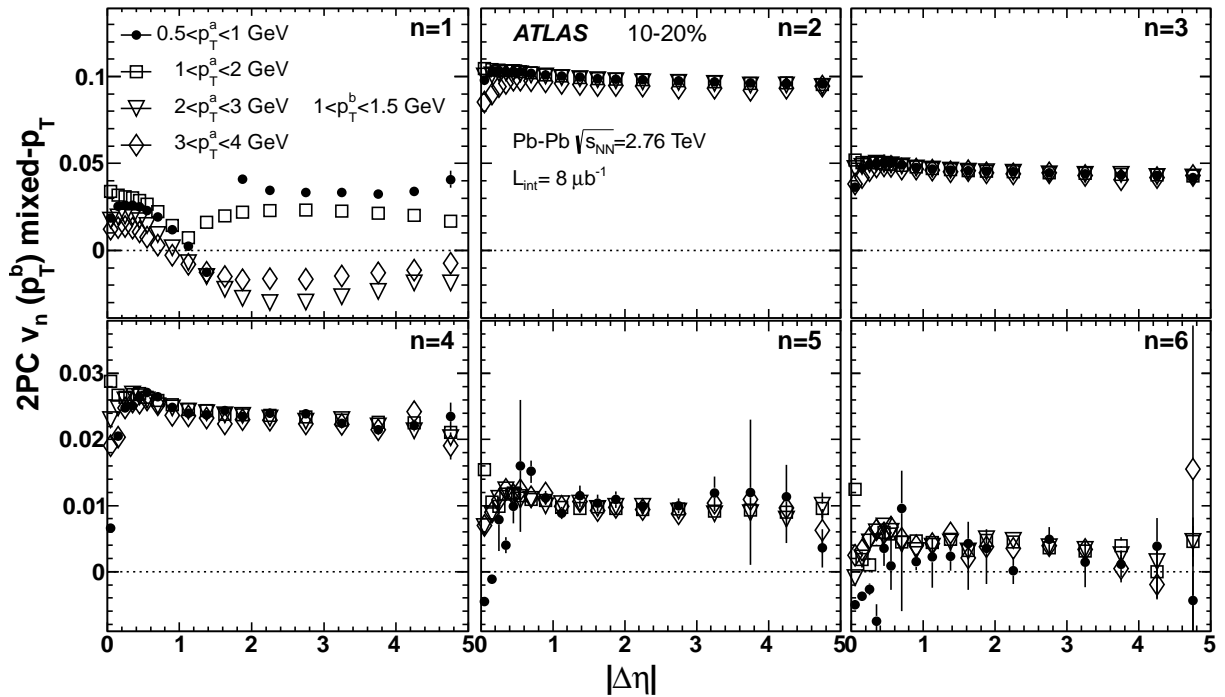}
\caption{\label{fig:refact1} The values of $v_{n}(\pT^{\mathrm b})=\frac{v_{n,n}(\pT^{\mathrm a},\pT^{\mathrm b})}{v_n(\pT^{\mathrm a})}$ vs. $|\Delta\eta=\eta_{\mathrm{a}}-\eta_{\mathrm{b}}|$ for $1<\pT^{\mathrm b}<1.5$ GeV, calculated from a reference $v_n(\pT^{\mathrm a})$ in four $\pT^{\mathrm a}$ ranges (0.5--1, 1--2, 2--3, and 3--4 GeV)~\cite{Aad:2012bu}. The error bars indicate the statistical uncertainties.}
\end{figure}

During the Quark Matter 2011 conference, all three LHC experiments demonstrated that the non-flow effects are small for $n=2-6$ and the factorization:
\begin{eqnarray}
\label{eq:4}
v_{n,n}(\pT^{\mathrm a},\pT^{\mathrm b})  = v_n(\pT^{\mathrm a})v_n(\pT^{\mathrm b})
\end{eqnarray}
is valid within $\sim 5-10\%$ (see S. Mohapatra's talk in this workshop~\cite{soumya}). They further show that this factorization is valid as long as one of the particle has $\pT\lesssim4$ GeV, beyond that the influence of dijet correlation becomes large. In contrast, the factorization breaks down for $n=1$ over the entire $\pT$ and $\eta$ range. This is illustrated in Figure~\ref{fig:refact1}, which shows the $v_n$ values calculated from $v_{n,n}$ via Eq.~\ref{eq:4}, assuming no non-flow effect; They are obtained for $1<\pT^{\mathrm b}<1.5$ GeV using four different reference $\pT^{\mathrm a}$ ranges. The extracted $v_2-v_6$ are independent of reference $\pT^{\mathrm a}$ for $|\Delta\eta|>1$, while the $v_1(\pT^{\mathrm b})$ values calculated this way clearly change with the choice of $\pT^{\mathrm a}$, indicating a breakdown of the factorization over the measured $\pT$ and $\Delta\eta$ range. 

The main reason for the lack of factorization for $v_{1,1}$ is that it is strongly influenced by GMC, while all higher-order coefficients conserve momentum due to their multi-fold symmetries. One example of GMC effects comes from dijet fragmentation in peripheral events, which tend to give a negative $v_{1,1}$ at large $|\Delta\eta|$. The influence of GMC on the two-particle correlation was studied extensively~\cite{Borghini:2000cm,Gardim:2011qn}, and was shown to explicitly break the factorization relation Eq.~\ref{eq:4}~\cite{Borghini:2000cm}:
\begin{eqnarray}
\nonumber
v_{1,1}(\pT^{\mathrm a},\pT^{\mathrm b},\eta_{\mathrm a},\eta_{\mathrm b}) \approx v_1(\pT^{\mathrm a},\eta_{\mathrm a})v_1(\pT^{\mathrm b},\eta_{\mathrm b})-\frac{\pT^{\mathrm a}\pT^{\mathrm b}}{M\langle \pT^2\rangle}\;,\\\label{eq:v0}
\end{eqnarray}
where $M$ and $\langle \pT^2\rangle$ are the multiplicity and average squared transverse momentum for the whole event, respectively. The second term is a leading-order approximation for the GMC. This approximation is expected to be valid when the correction term is much smaller than one. The GMC effect is important in peripheral events and for high $\pT$ particles, but is diluted in central events due to the large multiplicity. 

The dependence of the $v_1$ on $\eta$ can be generally separated into a rapidity-odd component and a rapidity-even component. The rapidity-odd $v_1$ is thought to arise from the deflection of the colliding ions~\cite{Voloshin:2008dg}, and changes sign from negative $\eta$ to positive $\eta$. However, the rapidity-odd $v_1$ is generally small $<0.005$ for $|\eta|<2$ at the RHIC and LHC. The corresponding contribution to $v_{1,1}$ ($<2.5\times10^{-5}$) is negligible compared to the typical $v_{1,1}$ values (see below). The rapidity-even $v_1$ signal is argued to be associated to the dipole anisotropy of the pressure gradient arising from the dipole asymmetry of the nuclear overlap; this dipole asymmetry is mainly due to fluctuations in the initial geometry~\cite{Teaney:2010vd}. If the rapidity-even $v_1$ depends weakly on $\eta$, similar to the higher-order $v_n$, Eq.~\ref{eq:v0} can be simplified to:
\begin{eqnarray}
\label{eq:v1}
v_{1,1}(\pT^{\mathrm a},\pT^{\mathrm b}) \approx v_1(\pT^{\mathrm a})v_1(\pT^{\mathrm b})-\frac{\pT^{\mathrm a}\pT^{\mathrm b}}{M\langle \pT^2\rangle}\;.
\end{eqnarray}

\begin{figure}[!t]
\includegraphics[width=0.92\linewidth]{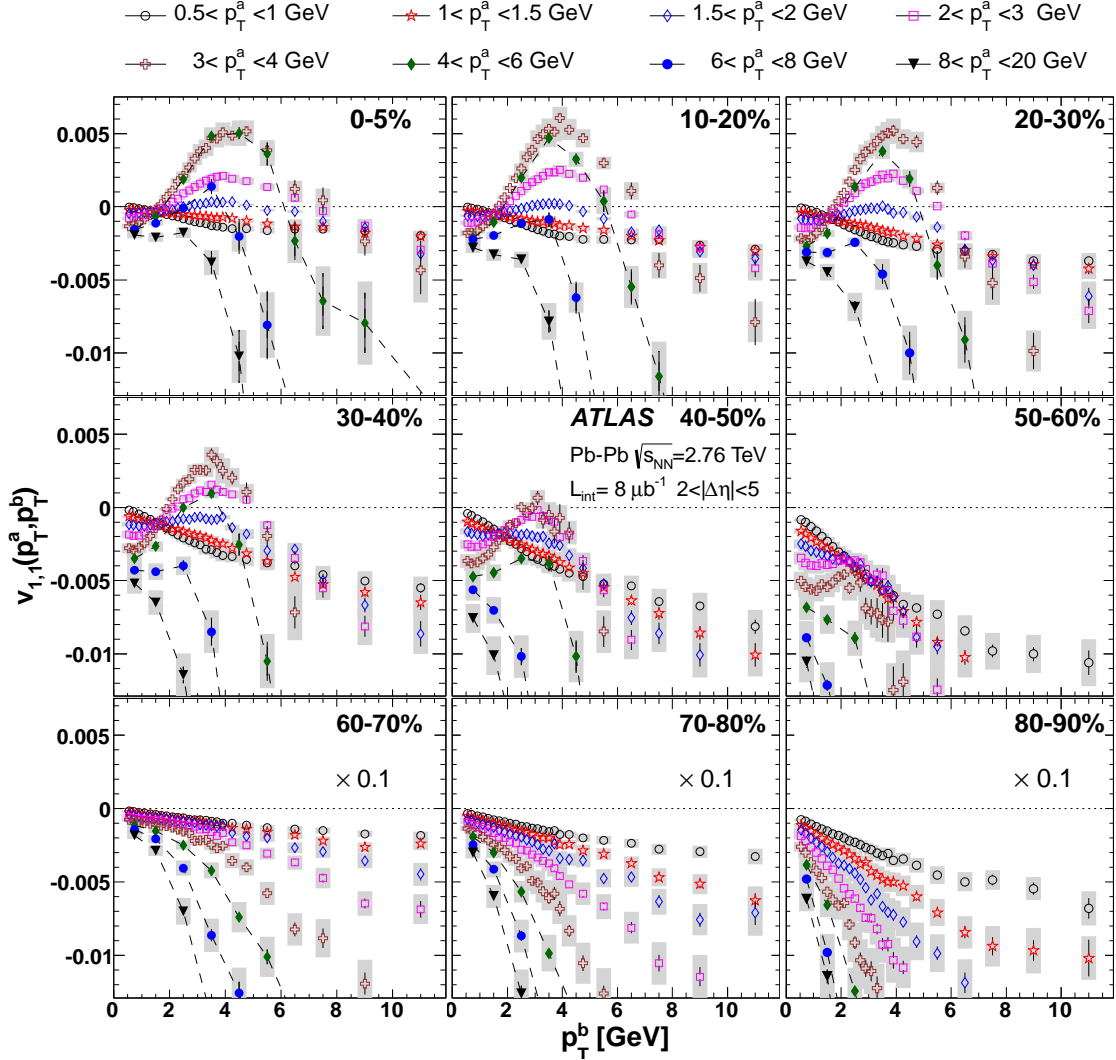}
\caption{\label{fig:v1fac3} The values of $v_{1,1}(\pT^{\mathrm a},\pT^{\mathrm b})$ for $2<|\Delta\eta|<5$ vs. $\pT^{\mathrm b}$ for different $\pT^{\mathrm a}$ ranges. Each panel presents results in one centrality interval. The error bars and shaded bands represent statistical and systematic uncertainties, respectively. The data points for the three highest $\pT^{\mathrm a}$ intervals have coarser binning in $\pT^{\mathrm b}$, hence are connected by dashed lines to guide the eye. The data in the bottom three panels are scaled down by a factor of ten to fit within the same vertical range.}
\end{figure}

The top-left panel of Figure~\ref{fig:refact1} indeed suggests that the $v_{1,1}$ values are independent of $\Delta\eta$ for $|\Delta\eta|>2$. Hence the $v_{1,1}$ values are integrated over $2<|\Delta\eta|<5$ to obtain one value for each $\pT^{\mathrm a}$ and $\pT^{\mathrm b}$ combination, and the resulting $v_{1,1}(\pT^{\mathrm a},\pT^{\mathrm b})$ functions are shown in Figure~\ref{fig:v1fac3}.  In peripheral events ($>40\%$ centrality), the $v_{1,1}$ values are always negative and their magnitudes increase nearly linearly with $\pT^{\mathrm a}$ and $\pT^{\mathrm b}$, due to the contribution from the away-side jet. In more central events, the contribution of this negative $v_{1,1}$ component is smaller, reflecting a dilution of the GMC term by the large event multiplicity. Furthermore, Figure~\ref{fig:v1fac3} clearly suggests that a positive $v_{1,1}$ component sets in for $2\lesssim\pT\lesssim6$ GeV. Its magnitude increases with $\pT$ and eventually drives $v_{1,1}$ into the positive region.

To extract the rapidity-even $v_1$, a least-squares fit of the $v_{1,1}$ data is performed for each centrality interval with the following form:
\begin{eqnarray}
\nonumber\chi^2 = \sum_{\mathrm {a,b}}\frac{\left(v_{1,1}(\pT^{\mathrm a},\pT^{\mathrm b}) - [v_1^{\mathrm{Fit}}(\pT^{\mathrm a})v_1^{\mathrm{Fit}}(\pT^{\mathrm b})-c\pT^{\mathrm a}\pT^{\mathrm b}]\right)^2}{\left(\sigma_{\mathrm {a,b}}^{\mathrm{stat}}\right)^2+\left(\sigma_{\mathrm {a,b}}^{\mathrm{sys,p2p}}\right)^2}\;,\\\label{eq:v2}
\end{eqnarray}
where $\sigma_{\mathrm {a,b}}^{\mathrm {stat}}$ and $\sigma_{\mathrm {a,b}}^{\mathrm {sys,p2p}}$ denote the statistical and point-to-point systematic uncertainties for $v_{1,1}(\pT^{\mathrm a},\pT^{\mathrm b})$, respectively. The $v_1^{\mathrm{Fit}}(\pT)$ function is defined via a smooth interpolation of its values at 15 discrete $\pT$ points $v_1^{\mathrm{Fit}}(p_{\mathrm T,i})|_{i=1}^{15}$ (see Table~\ref{tab:fit}), and these together with the parameter $c$ constitute a total of 16 fit parameters. This interpolation procedure is motivated by the fact that the function $v_1^{\mathrm{Fit}}(\pT)$ is unknown but is expected to vary smoothly with $\pT$. In the peripheral collisions where the $v_1$ contribution is small, the fit is also repeated including only the GMC contribution (i.e. removing the $v_1^{\mathrm{Fit}}(\pT^{\mathrm a})v_1^{\mathrm{Fit}}(\pT^{\mathrm b})$ in Eq.~\ref{eq:v2}). 

\begin{figure}[!t]
\includegraphics[width=1\linewidth]{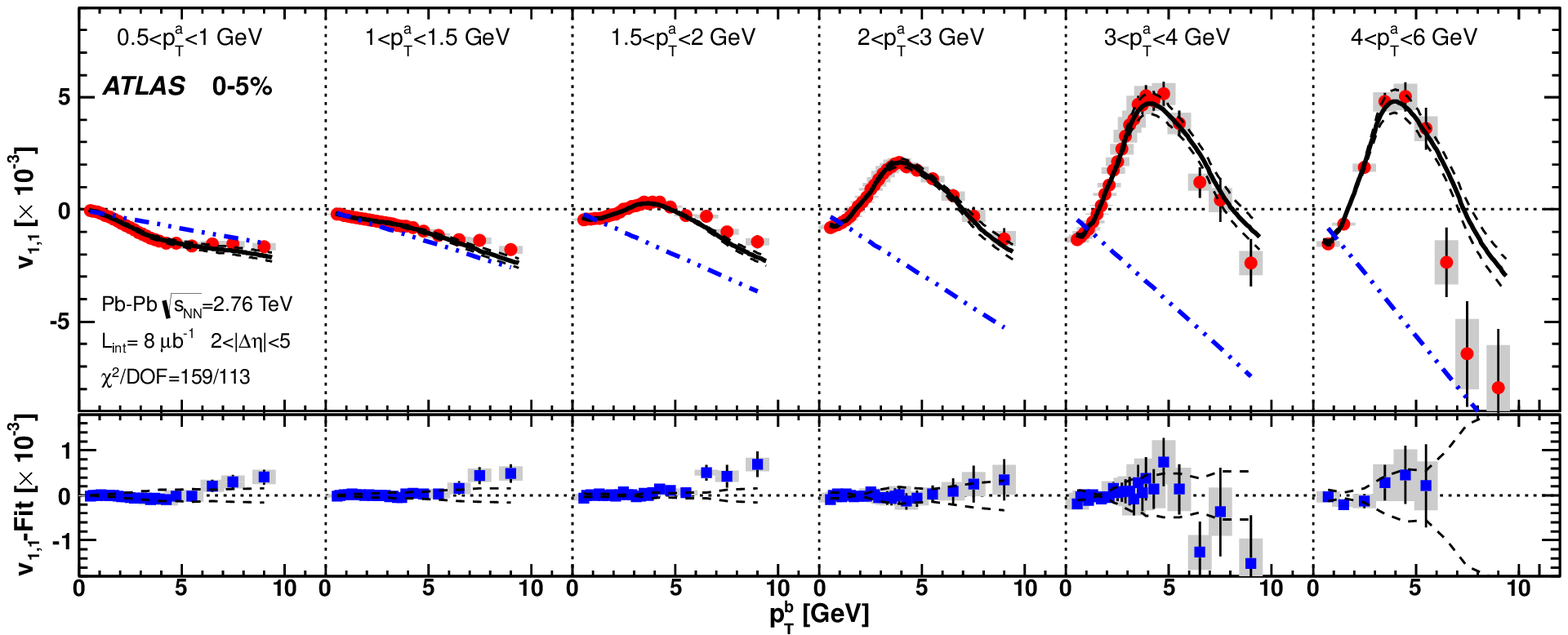}\\
\includegraphics[width=1\linewidth]{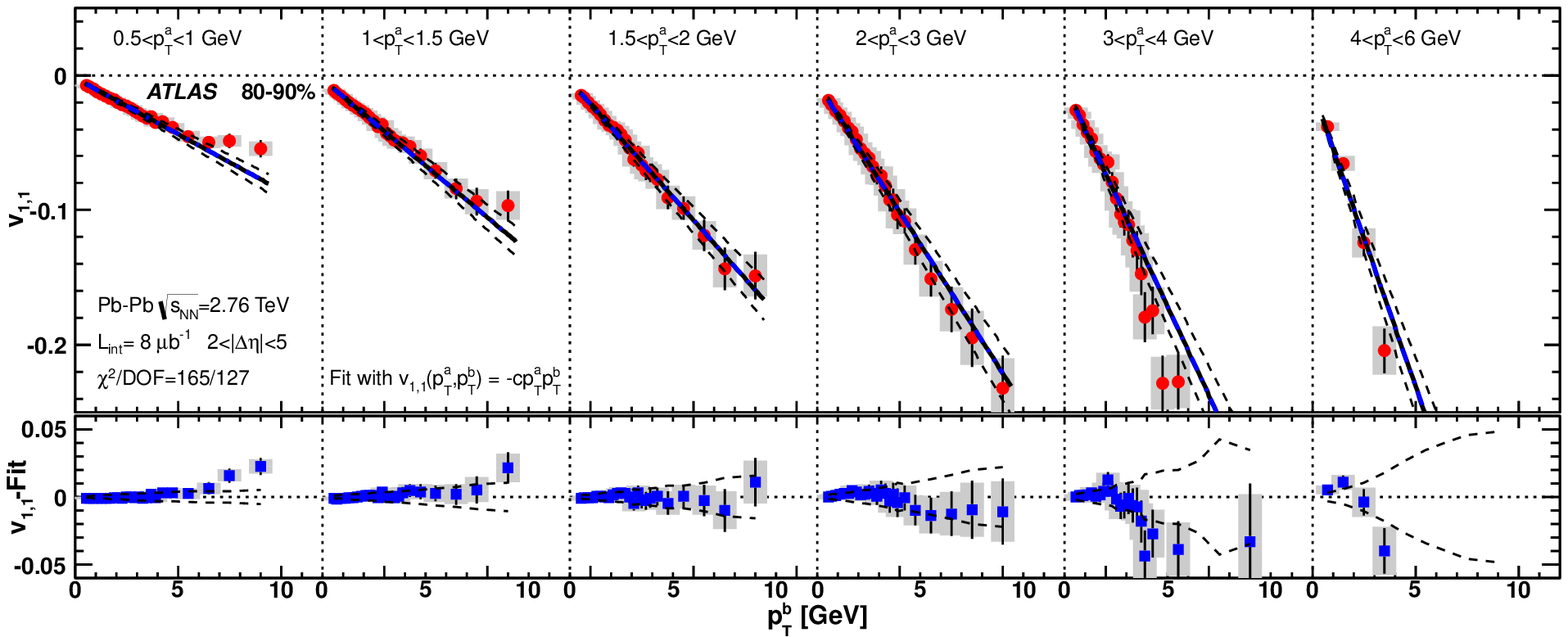}
\caption{\label{fig:v1fit1} Top panel: Global fit to the $v_{1,1}$ data for the (0--5)\% centrality interval via Eq.~\ref{eq:v2}. The fit is performed simultaneously over all $v_{1,1}$ data points in a given centrality interval, organized as a function of $\pT^{\mathrm b}$ for various $\pT^{\mathrm a}$ ranges, with shaded bars indicating the correlated systematic uncertainties. The fit function and the associated systematic uncertainties are indicated by the thick-solid lines and surrounding dashed lines, respectively. The dot-dashed lines intercepting each dataset (with negative slope) indicate estimated contributions from the momentum conservation component. The lower part of the panel shows the difference between the data and fit (solid points), as well as the systematic uncertainties of the fit (dashed lines). Bottom panel: Global fit for the (80--90)\% centrality interval considering only the GMC contribution.}
\end{figure}

The top panel of Figure~\ref{fig:v1fit1} shows the fit to the $v_{1,1}$ data from Figure~\ref{fig:v1fac3} for the (0--5)\% centrality interval. The $v_{1,1}$ data in Figure~\ref{fig:v1fac3} are plotted as a function of $\pT^{\mathrm b}$ for six intervals of $\pT^{\mathrm a}$. The seemingly-complex patterns of the $v_{1,1}$ data in (0-5)\% centrality interval are well described by the two-component fit across broad ranges of $\pT^{\mathrm a}$ and $\pT^{\mathrm b}$, with the dot-dashed lines indicating the estimated contributions from the GMC. The typical $\chi^2/$DOF of the fit is between one and two depending on the centrality, as shown in Table~\ref{tab:fit}. The deviations of the data from the fit, as shown in the bottom section of each panel, are less than $10^{-4}$ for $\pT<5$ GeV. Above that $\pT$ and in more peripheral events the deviations increase, possibly reflecting the limitation of the leading-order approximation for the momentum conservation term in Eq.~\ref{eq:v1}, or the two-component assumption in general. 

The bottom panel of Figure~\ref{fig:v1fit1} shows the fit to the $v_{1,1}$ data from Figure~\ref{fig:v1fac3} for the (80--90)\% centrality interval, including only the CMC term. The data are consistent with the GMC up to 10 GeV in both $\pT^{\mathrm a}$ and $\pT^{\mathrm b}$, well into the region where the contribution from away-side jet is expected to dominate the 2PC. This behavior suggests that in the very peripheral collisions, the system that conserves the momentum involves mainly the particles from dijets.

\begin{figure}[!t]
\includegraphics[width=1\linewidth]{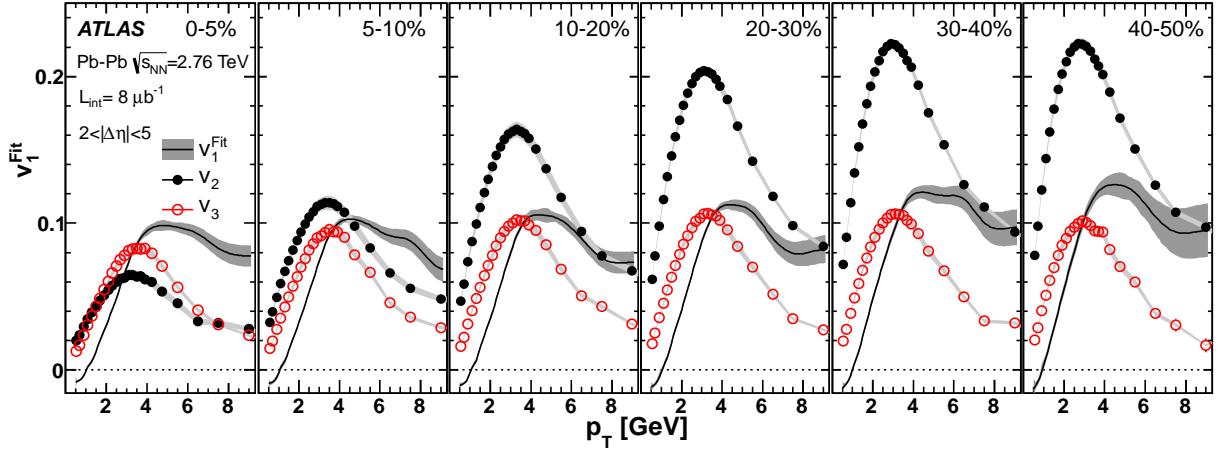}
\caption{\label{fig:v1fit2} $v_1^{\mathrm {Fit}}$ vs. $\pT$ for various centrality intervals. The shaded bands indicate the total uncertainty. The uncertainty bands are reproduced on their own at the bottom of the figure for clarity.}
\end{figure}

Figure~\ref{fig:v1fit2} shows the extracted $v_1^{\mathrm {Fit}}(\pT)$ for various centrality intervals together with the $v_2$ and $v_3$ data measured using the event plane method~\cite{Aad:2012bu}. The $v_1^{\mathrm{Fit}}(\pT)$ is negative for $\pT\lesssim1.0$ GeV (decrease to about 0.9 GeV in (40-50)\% centrality interval), confirming a generic feature expected for collective $v_1$ as suggested by hydrodynamic model calculations~\cite{Gardim:2011qn,Teaney:2010vd}. The crossing point extracted from ALICE data~\cite{Retinskaya:2012ky,Aamodt:2011by} is significantly higher ($\sim1.3-1.4$ GeV), which could be related to the fact the $|\Delta\eta|$ gap of 0.8 in ALICE data is not sufficient to fully suppress the near-side peak (see top left panel of Figure~\ref{fig:refact1} or Figure 2c of~\cite{Aad:2012bu}).  Nevertheless, that calculation clearly demonstrates that both the zero crossing point and the peak value are sensitive to the shear viscosity to entropy ratio, making our precision $v_1$ measurement a valuable tool for constraining the transport properties of the medium.

At higher $\pT$, the measured $v_1^{\mathrm{Fit}}$ values reach a maximum between 4 GeV and 5 GeV and then fall. The magnitude of the $v_1^{\mathrm{Fit}}$ is large: its peak value is comparable or slightly larger than that for the $v_3$. The peak value increases by about 20\% over the measured centrality range, also significantly larger than the increase for $v_3$ (about $\sim10\%$ over the same centrality range). The falloff of $v_1^{\mathrm{Fit}}(\pT)$ at higher $\pT$ may indicate the onset of path-length dependent jet energy loss, which correlates with the dipole asymmetry in the initial geometry to produce the large $v_1$ coefficient at high $\pT$. This falloff is slower than that for the $v_2$ and $v_3$, with an magnitude that is surprisely larger than $v_2$ in central collisions, consistent with a finding based on jet absorption model calculation~\cite{Jia:2012ez}. These high $\pT$ behaviors suggest that our $v_1^{\mathrm{Fit}}(\pT)$ results also provide valuable input for understanding jet quenching phenomena.

\begin{figure}[!t]
\begin{tabular}{lr}
\begin{minipage}{0.62\linewidth}
\begin{flushleft}
\includegraphics[width=1\linewidth]{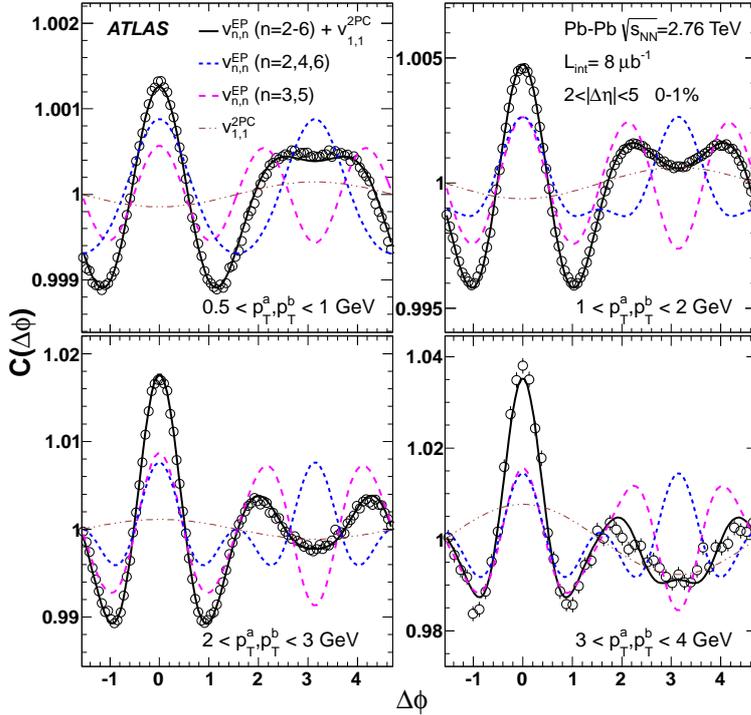}
\end{flushleft}
\end{minipage}
\begin{minipage}{0.35\linewidth}
\begin{flushright}
\caption{\label{fig:rec0b} Measured correlation functions compared with those reconstructed from $v_2$--$v_6$ measured from the FCal$_{\mathrm{P(N)}}$ method and $v_{1,1}$ from the 2PC method in (0--1)\% centrality interval for four $\pT$ ranges. The contributions from $n=1$, $n=3,5$ and $n=2,4,6$ are shown separately. The error bars indicate the statistical uncertainties.}
\end{flushright}
\end{minipage}
\end{tabular}
\end{figure}

Finally, results in Figure~\ref{fig:v1fit2} also imply that the rapidity-even collective $v_1$ is an important component of the two-particle correlation at intermediate $\pT$. For example, the large positive $v_{1,1}$ harmonic in Figure~\ref{fig:rec0b}, responsible for the mach-cone like structure in the the bottom-right panel, is mainly due to the rapidity-even $v_1$: its contribution to $v_{1,1}$ can be three times larger than the negative momentum conservation term estimated by the global fit (top panel of Figure~\ref{fig:v1fit1}). 

If the two-component ansatz of Eq.~\ref{eq:v1} is valid, the fit parameter $c$ should be inversely proportional to multiplicity $M$ and $\langle\pT^2\rangle$ of the whole event. This ansatz is checked by calculating the product of $c$ and the charged hadron multiplicity at mid-rapidity $\frac{dN}{d\eta}_{|\eta=0}$ from~\cite{ATLAS:2011ag}, assuming $\frac{dN}{d\eta}_{|\eta=0}\propto M$. The results are summarized in Table~\ref{tab:fit} for each centrality interval. Since $\langle\pT^2\rangle$ for the whole event is expected to vary weakly with centrality, the product is also expected to vary weakly with centrality. Table~\ref{tab:fit} shows that this is indeed the case, supporting the assumptions underlying Eq.~\ref{eq:v1}.

It has been argued that the system that conserves momentum may only involves a subset of the event~\cite{Chajecki:2008yi}. If the value of $\langle\pT^2\rangle$ is known, one can estimate the effective size of the system that is correlated due to the GMC. For $\langle\pT^2\rangle\sim1$ GeV$^2$ (estimated from charged hadron spectrum from ALICE~\cite{Preghenella:2011vy}), the system size is estimated to be $M\sim5000$ in (0-5)\% most central collisions, covering about 3 units in $\eta$; the system size increases to about 4 units for (40-50)\% centrality interval.

\begin{table}[!h]
\begin{center}
\begin{tabular}{l c c c }\tabularnewline\hline
Centrality        & $\chi^2$/DOF& $c$[$\times$0.001GeV$^{-2}$]  & $c\frac{dN}{d\eta}_{|\eta=0}$[$\times$GeV$^{-2}$]\tabularnewline\hline
   0-5\%          & 159/113     & $0.24\pm0.02$   & $0.39\pm 0.04$\tabularnewline\hline
   5-10\%         & 133/113     & $0.28\pm0.02$   & $0.37\pm 0.04$\tabularnewline\hline
   10-20\%        & 165/113     & $0.35\pm0.03$   & $0.36\pm 0.04$\tabularnewline\hline
   20-30\%        & 134/113     & $0.50\pm0.04$   & $0.34\pm 0.03$\tabularnewline\hline
   30-40\%        & 188/113     & $0.75\pm0.05$   & $0.33\pm 0.03$ \tabularnewline\hline
   40-50\%        & 181/113     & $1.16\pm0.09$   & $0.32\pm 0.03$\tabularnewline\hline\hline
 \multicolumn{4}{c}{15 interpolation points used in the default fit:}  \tabularnewline
 \multicolumn{4}{c}{0.5, 0.7, 0.9, 1.1, 1.3, 1.5, 2.0, 2.5, 3.0, 3.5, 4.5, 5.5, 6.5, 7.5, 9.0 GeV}\tabularnewline\hline
\end{tabular}
\end{center}
\caption{\label{tab:fit} Quality of the fit $\chi^2$/DOF, fit parameter $c$, and corresponding multiplicity scaled values $c\frac{dN}{d\eta}_{|\eta=0}$ for various centrality intervals. The uncertainty of $c\frac{dN}{d\eta}_{|\eta=0}$ is calculated as the quadrature sum of uncertainties from $c$ and $\frac{dN}{d\eta}_{|\eta=0}$ of Ref.~\cite{ATLAS:2011ag}. The bottom row lists the 15 $\pT$ interpolation points used in the default fit.}
\end{table}

\begin{figure}[!t]
\includegraphics[width= 0.49\linewidth]{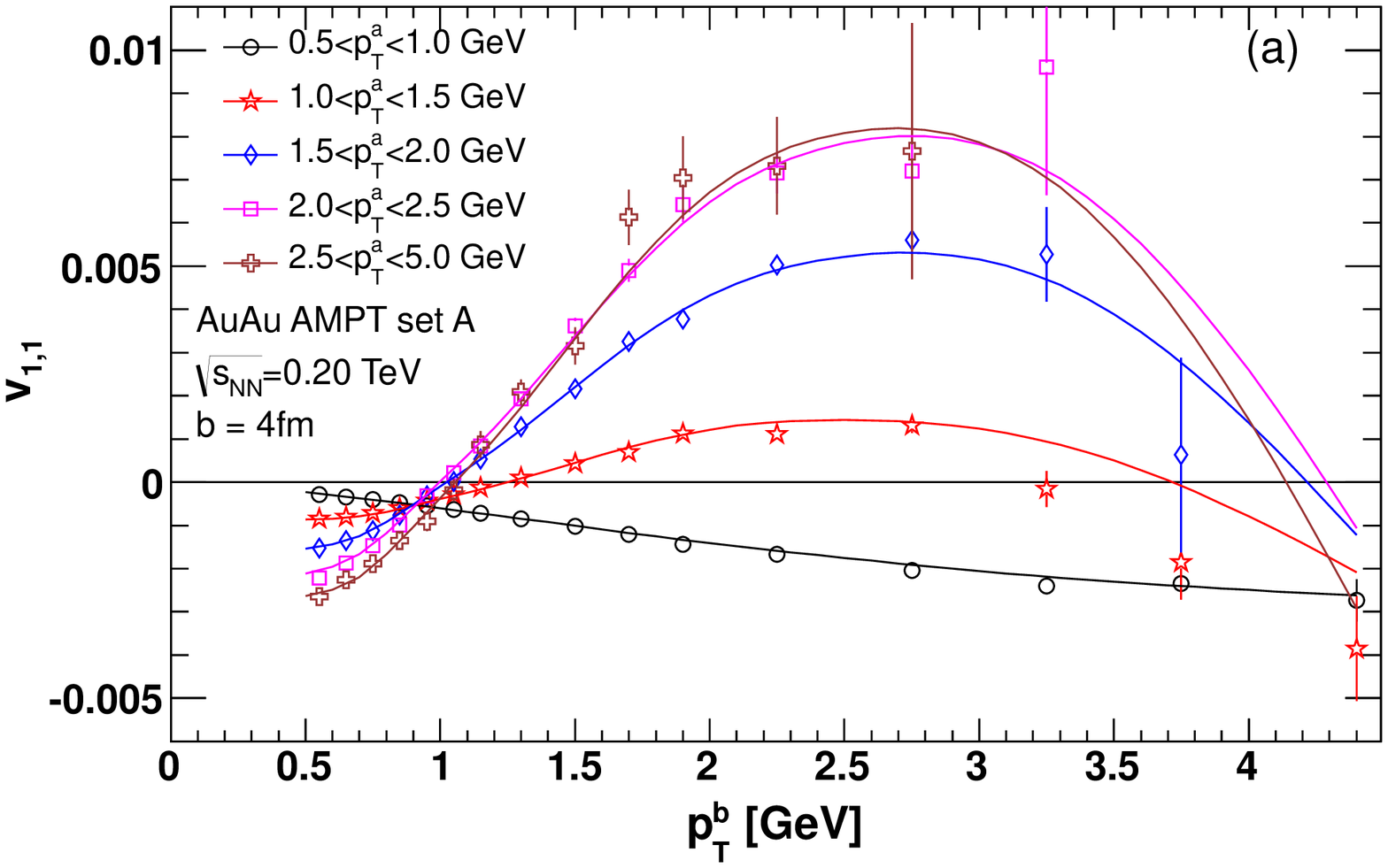}\includegraphics[width= 0.49\linewidth]{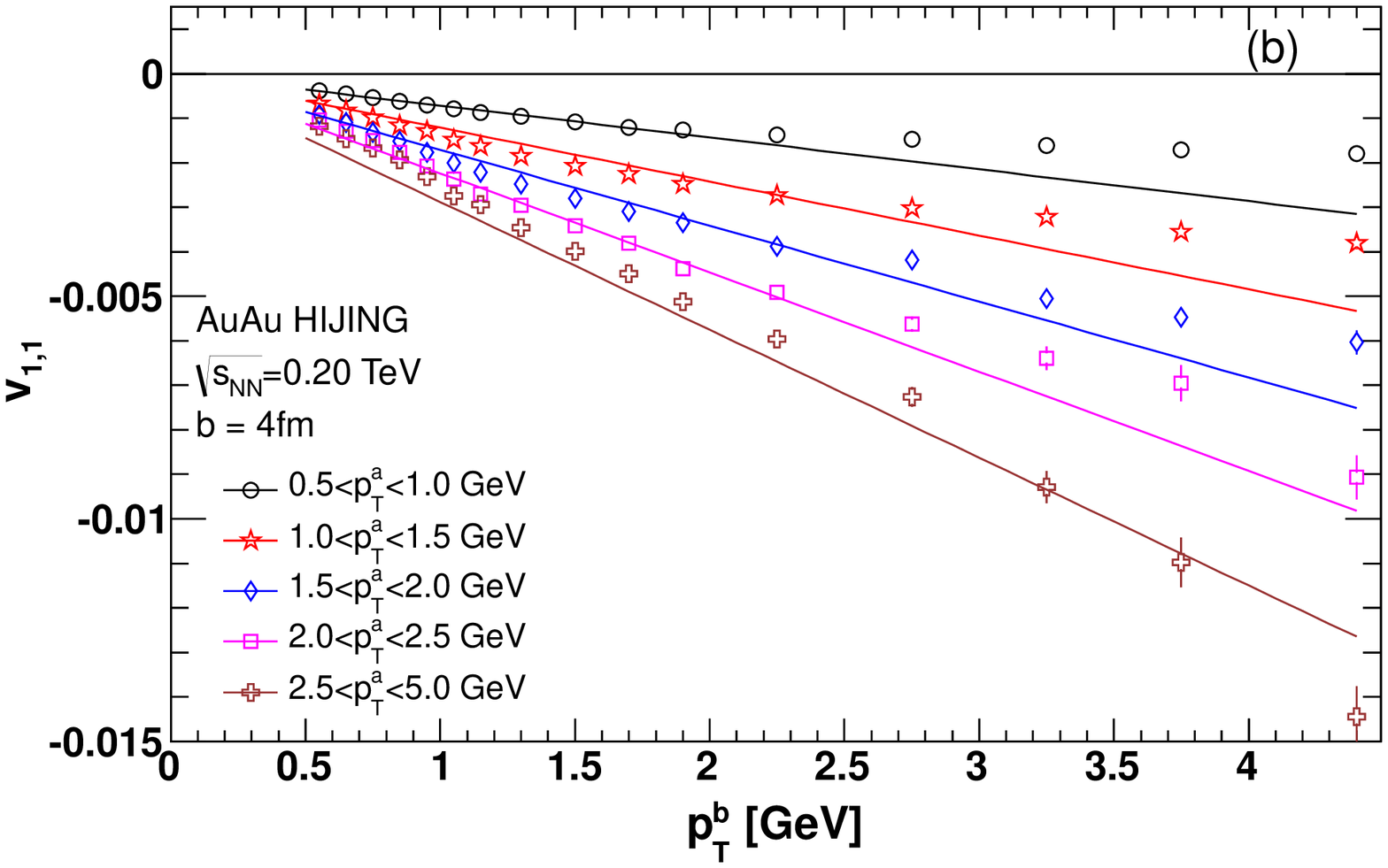}
\includegraphics[width= 0.49\linewidth]{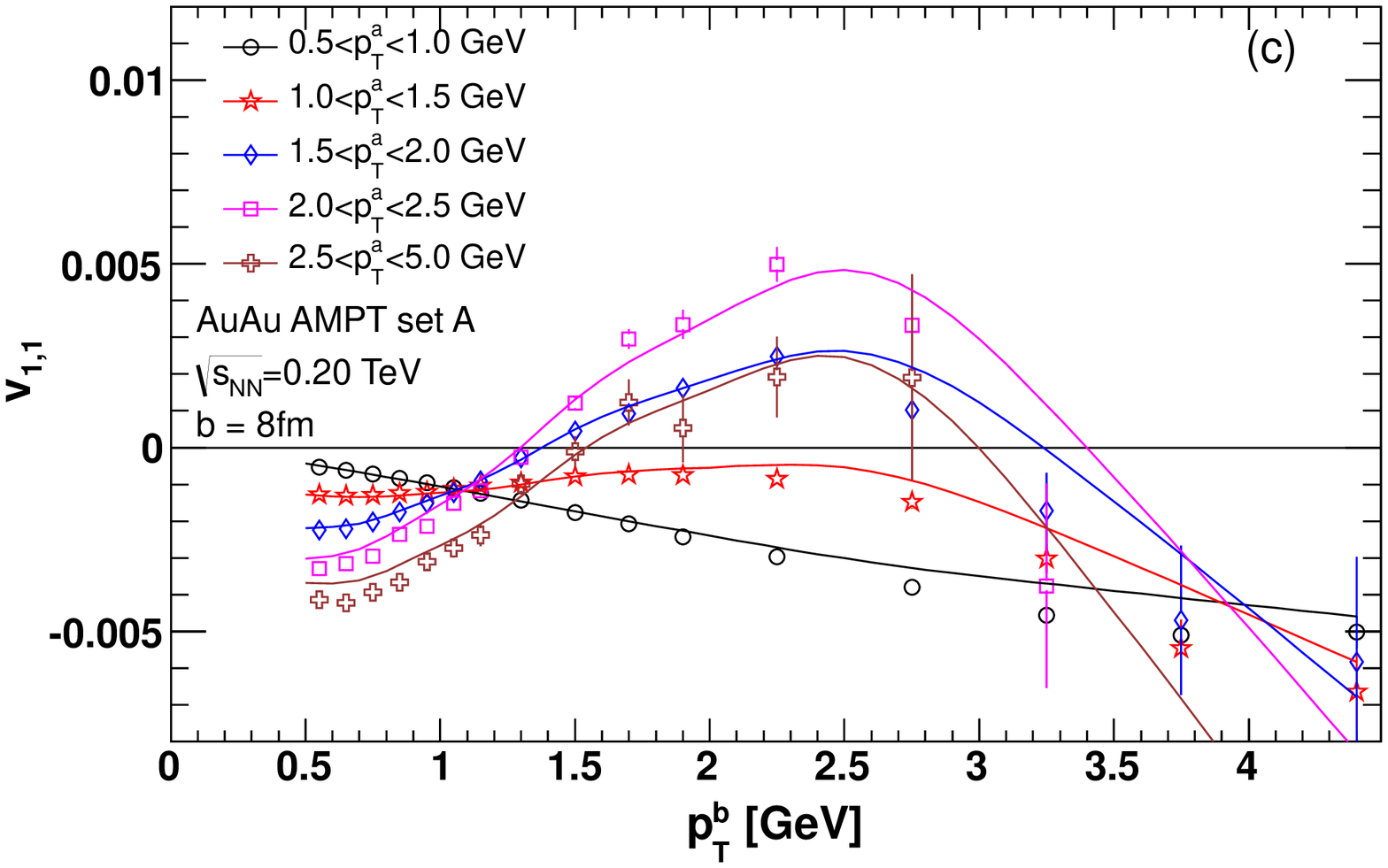}\includegraphics[width= 0.49\linewidth]{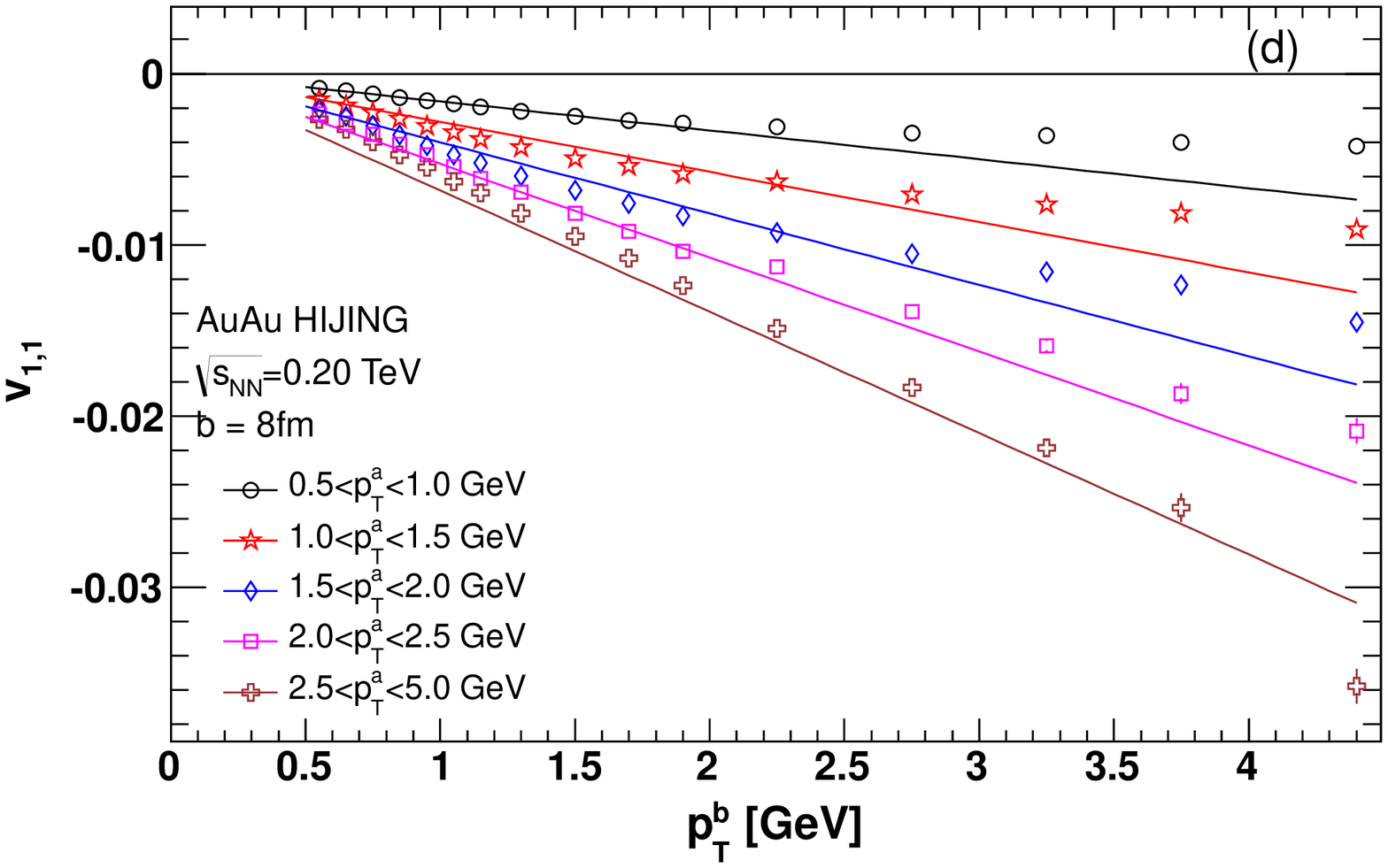}
\caption{\label{fig:1}  $v_{1,1}(\pT^{\mathrm a},\pT^{\mathrm b})$ (symbols) from AMPT (left panels) and HIJING (right panels) for $b=4$~fm (top panels) and $b=8$~fm in AuAu collisions at $\sqrt{s_{NN}}=0.2$~TeV. Lines in the left (right) panels are result of a global fit to Eq.~\ref{eq:v1} (a fit to pure momentum conservation component: $c\pT^{\mathrm a}\pT^{\mathrm b}$). Taken from~\cite{Jia:2012gu}.}
\end{figure}

Equation~\ref{eq:v1} was derived based on fairly general grounds, hence it is expected to be valid for any Monte Carlo model that contains the flow physics and respects momentum conservation. One example is the AMPT (A Multi Phase Transport) model~\cite{Lin:2004en}. This model combines the initial condition from HIJING and final state interaction via a parton and hadron transport model. The initial condition from HIJING is seeded by a Glauber model for nucleon-nucleon scatterings, hence it contains the fluctuations which generate event-by-event spatial asymmetries; the parton and hadron transport is responsible for transforming the initial asymmetries into the momentum anisotropy. Figure~\ref{fig:1} shows the factorization behavior from both HIJING and AMPT at the RHIC and LHC energies obtained from~\cite{Jia:2012gu}. The $v_{1,1}(\pT^{\mathrm a},\pT^{\mathrm b})$ values from HIJING model, which has no final state interactions, exhibit linear dependence on $\pT^{\mathrm a}$, $\pT^{\mathrm b}$, consistent with correlations coming from only GMC. However once the final state interactions are switched on as is in the AMPT model, the $v_{1,1}(\pT^{\mathrm a},\pT^{\mathrm b})$ data develops crossing pattern at $\pT\sim1$ GeV and hump-back structures in $2<\pT<6$ GeV ranges, very similar to what is observed in the data. A simple two-component fit via Eq.~\ref{eq:v1} is able to describe these structures simultaneously. This clearly demonstrates the first component of Eq.~\ref{eq:v1} is indeed associated with the dipolar flow arising from the initial geometry fluctuations and strong interactions in the final state.

In summary, the $v_{1,1}=\langle \cos(\Delta\phi)\rangle$ component of the two-particle correlation data are extracted for Pb-Pb collisions at $\sqrt{s_{_{\mathrm{NN}}}}=2.76$ TeV. The $\pT$ and $\Delta\eta$ dependence of $v_{1,1}$ data are found to be consistent with contributions from rapidity-even dipolar flow $v_1$ and global momentum conservation. A two-component fit is used to extract the individual contributions from these two components. The extracted $v_1$ function crosses zero at $\pT\sim1$ GeV, reaches a value of 0.1-0.12 (comparable or slightly larger than $v_3$), and then decreases at higher $\pT$. The $\pT$ at which $v_1$ reaches maximum is about 1 GeV higher than other $v_n$. Extracted $v_1$ shows a mild but stronger increase with centrality ($\sim20\%$) than the $v_3$. The magnitude of the extracted momentum conservation component suggests that the system conserving momentum involves only a subset of the event (spanning about 3 units in $\eta$ in central collisions). Simulation based on AMPT model qualitatively reproduces the complex $\pT$ dependence of the $v_{1,1}$. Together with HIJING model simulation, they supports the two-component interpretation of the $v_{1,1}$ data and strongly suggest that the presence of dipolar flow requires both initial geometry fluctuations and strong final state interactions.

This work is in part supported by NSF under award number PHY-1019387.

\end{document}